\documentclass[12pt]{iopart}
\usepackage{iopams}
\usepackage{epsfig}
\def\noi{\noindent}

\def\bea{\begin{eqnarray}}  \def\eea{\end{eqnarray}}
\def\beq{\begin{equation}}   \def\eeq{\end{equation}}

\def\beeq{\begin{eqnarray}} \def\eeeq{\end{eqnarray}}

\begin{document}

\title[$\pi^0$ Fixed p$_{\bot}$ suppression and elliptic flow at LHC]{$\pi^0$
Fixed p$_{\bot}$ suppression and elliptic flow at LHC}

\author{A. Capella$^1$, E. G. Ferreiro$^2$, A. Kaidalov$^3$ and K. Tywoniuk$^4$}

\address{$^1$ 
Laboratoire de Physique Th\'eorique,
Universit\'e de Paris XI, B\^atiment 210, 91405 Orsay Cedex, France}
\address{$^2$
Departamento de F{\'\i}sica de Part{\'\i}culas, 
Universidad de Santiago de Compostela, 15782 Santiago de Compostela, 
Spain}
\address{$^3$
ITEP, 117259 Moscow, Russia}
\address{$^4$
Department of Physics, University of Oslo, 0316 Oslo, Norway}
\begin{abstract}
Using a final state interaction model which describes the data on these two observables, at RHIC, we make predictions at the LHC -- using the same cross-section and $p_{\bot}$-shift. The increase in the medium density between these two energies (by a factor close to three) produces an increase of the fixed $p_{\bot}$ $\pi^0$ suppression by a factor 2 at large $p_{\bot}$ and of $v_2$ by a factor 1.5.

\end{abstract}

\vskip -0.5cm
\section{$\pi^0$ Fixed $p_{\bot}$ suppression} 
Final state interaction (FSI) effects have been observed in $AA$ collisions. They are responsible of strangeness enhancement, $J/\psi$ supression, fixed $p_{\bot}$ supression, azimuthal asymmetry, ... Is it the manifestation of the formation of a new state of matter or can it be described in a FSI model with no reference to an equation of state, thermalization, hydrodynamics, ...~? We take the latter view and try to describe all these obseervables within a unique formalism~: the well known gain and loss differential equations. We assume \cite{Capella:2004xj} that, at least for particles with $p_{\bot}$ larger than $<p_{\bot} >$, the interaction with the hot medium produces a $p_{\bot}$-shift $\delta p_{\bot}$ towards lower values and thus the yield at a given $p_{\bot}$ is reduced. There is also a gain term due to particles produced at $p_{\bot} + \delta p_{\bot}$. Due to the strong decrease of the $p_{\bot}$-distributions with increasing $p_{\bot}$, the loss is much larger than the gain. Asuming boost invariance and dilution of the densities in $1/\tau$ due to longitudinal expansion, we obtain 
\beq
\label{1e}
{\tau dN_{\pi^0} (b, s, p_{\bot}) \over d\tau}= - \sigma N(b,s) \left [ N_{\pi^0}(b,s,p_{\bot} ) - N_{\pi^0}(b,s,p_{\bot} + \delta p_{\bot})\right ]
\eeq

\noi Here $N \equiv dN/dy d^2s$ is the transverse density of the medium and $N_{\pi^0}$ the corresponding one of the $\pi^0$ \cite{Capella:2007bw}. This has to be integrated between initial time $\tau_0$ and freeze-out time $\tau_f$. The solution deepnds only on $\tau_f/\tau_0$. We use $\sigma = 1.4$~mb at both energies and $\delta p_{\bot} = p_{\bot}^{1.5}/20$ for $p_{\bot} < 2.9$~GeV and $\delta p_{\bot} = p_{\bot}^{0.8}/9.5$ for $p_{\bot} > 2.9$~GeV \cite{Capella:2006fw}. Eq. (\ref{1e}) at small $\tau$ describes an interaction at the partonic level. Indeed, here the densities are very large and the hadrons not yet formed. At later times the interaction is hadronic. Most of the effect takes place in the partonic phase. We use a single (effective) value of $\sigma$ for all values of the proper time $\tau$. 
The results at RHIC and LHC are given in Fig.~1. At LHC only shadowing \cite{2r} has been included in the initial state. The suppression is given by the dashed line. It coincides with $R_{AA}$ for $p_{\bot}$ large enough -- when shadowing and Cronin efffects are no longer present. The LHC suppression is thus a factor of two larger than at RHIC.

\begin{center}
\begin{figure*}
\begin{minipage}[t]{80mm}
\epsfxsize=7.5cm
\epsfysize=7.45cm
\centerline{\epsfbox{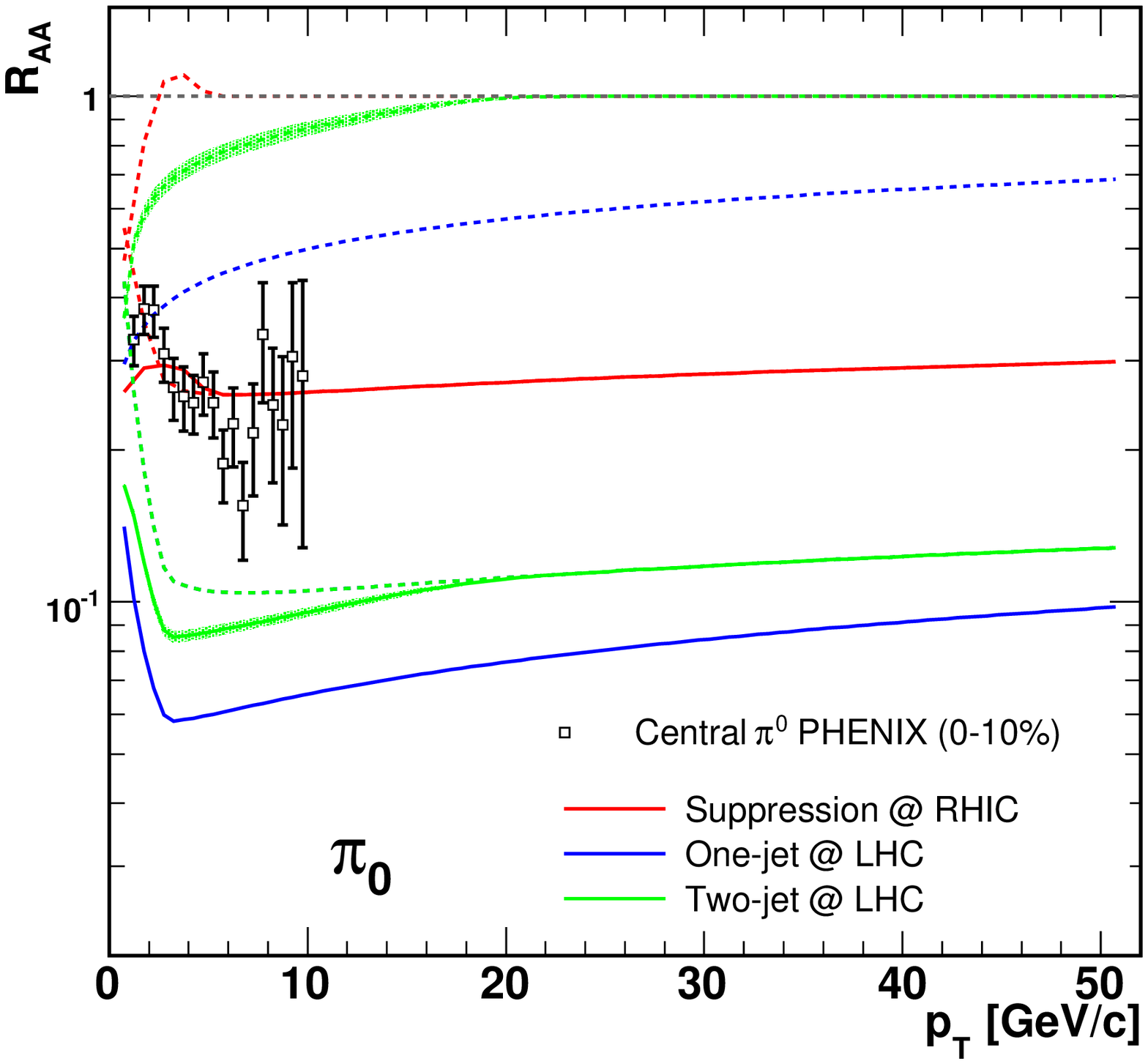}}
\vskip -0.25cm
\caption{
From up to down: RHIC initial, 2 LHC initial, 
RHIC final,
LHC FSI, LHC FSI+shadowing.} 
\end{minipage}
\hspace{\fill}
\begin{minipage}[t]{80mm}
\vskip -7.45cm
\epsfxsize=7.5cm
\epsfysize=7.45cm
\centerline{\epsfbox{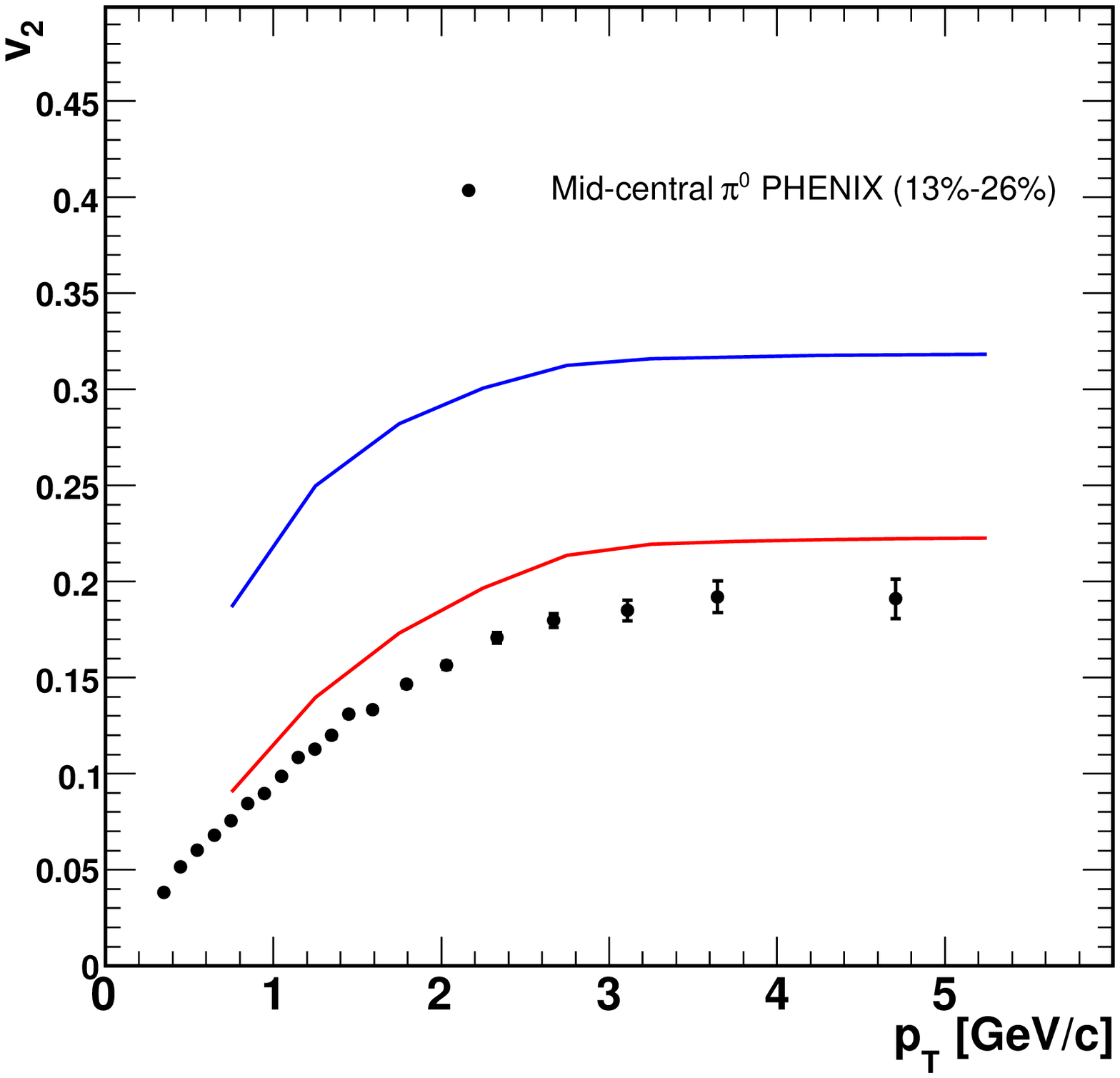}}
\vskip -0.25cm
\caption{$v_2$ 
for $\pi^0$ 
at 
RHIC (lower curve) and LHC (upper curve).}
\end{minipage}
\vskip -0.3cm
\end{figure*}
\end{center}

\section{Elliptic flow} 
Final state interaction in our approach gives rise to a positive $v_2$ \cite{Capella:2006fw} (no need for an equation of state or hydro). Indeed, when the $\pi^0$ is emitted at $\theta_R = 90^{\circ}$ its path length is maximal (maximal absorption). In order to compute it we assume that the density of the hot medium is proportional to the path length $R_{\theta_R}(b,s)$ of the $\pi^0$ inside the interaction region determined by its transverse position $s$ and its azimuthal angle $\theta_R$. Hence, we replace $N(b,s)$ by $N(b,s) R_{\theta_R}(b,s)/$\break\noindent $<R_{\theta_R}(b,s)>$ where $R_{\theta_R}$ is the $\pi^0$ path length and $<>$ denotes its average over $\theta_R$. (In this way the averaged transverse density $N(b,s)$ is unchanged). The suppression $S_{\pi^0}(b,s)$ depends now on $\theta_R$ and $v_2$ is given by
\beq
\label{2e}
v_2(b, p_{\bot}) = 
{\displaystyle{\int} d\theta_R S_{\pi^0} (b, p_{\bot}, \theta_R) 
\cos 2 \theta_R \over \displaystyle{\int} d\theta_R S_{\pi^0} (b, p_{\bot}, \theta_R)}
\eeq
%
The results at RHIC and LHC are presented in Fig. 2.

\end{document}